\documentstyle[12pt]{article}

\newcommand{\resection}[1]{\setcounter{equation}{0}\section{#1}}

\newcommand{\EQ}{\begin{equation}}
\newcommand{\EN}{\end{equation}}
\newcommand{\bea}{\begin{eqnarray}}
\newcommand{\eea}{\end{eqnarray}}
\newcommand{\hs}{\hspace{0.1cm}}

\begin{document}
\setcounter{page}{0}
\topmargin 0pt
\oddsidemargin 5mm
\renewcommand{\thefootnote}{\arabic{footnote}}
\newpage
\setcounter{page}{0}
\begin{titlepage}
\begin{flushright}
ISAS/EP/94/30
\end{flushright}
\vspace{0.5cm}
\begin{center}
{\large {\bf Statistical Models with a Line of Defect}}\\
\vspace{1.8cm}
{\large G. Delfino, G. Mussardo and P. Simonetti}\\
\vspace{0.5cm}
{\em International School for Advanced Studies,\\
and \\
Istituto Nazionale di Fisica Nucleare\\
34014 Trieste, Italy}\\

\end{center}
\vspace{1.2cm}

\renewcommand{\thefootnote}{\arabic{footnote}}
\setcounter{footnote}{0}

\begin{abstract}
\noindent
The factorization condition for the scattering amplitudes of
an integrable model with a line of defect gives rise to a set of
Reflection-Transmission equations. The solutions of these equations
in the case of diagonal $S$-matrix in the bulk are only those with
$S =\pm 1$. The choice $S=-1$ corresponds to the Ising model. We compute
the transmission and reflection amplitudes relative to the interaction of
the Majorana fermion with the defect and we discuss their relevant features.
\end{abstract}

\vspace{.3cm}

\end{titlepage}

\newpage
\resection{Introduction}
The theory of two-dimensional integrable statistical models with a finite
correlation length can be elegantly formulated in terms of an ensemble of
particle excitations in bootstrap interaction. Originally proposed by
Zamolodchikov for systems defined in the infinite volume \cite{Zam} (see
also
%\cite{YL,FZ,ZamPol,ZN,GM}),
[2-6]), the bootstrap approach has been recently generalized to the case
of integrable systems with boundary
%\cite{Cherednik,Sklyanin,GZ,KF,Sazaki,FS,Ghoshal}.
[7-13]. Here, in addition to the usual scattering matrices of the particles
in the bulk, one needs to introduce the reflection amplitudes relative to
the interaction with the boundary.

In this letter we will discuss the properties of integrable 2-d statistical
models in the presence of a line of defect. These systems may
interpolate between a bulk or a boundary statistical behaviour
and their theoretical understanding has stimulated quite a large literature
(see for instance
%\cite{def1,def2,Bariev,McCoy,Kad,Brown,Hanke,Burk,Ko}).
[14-22]). We initially derive the Reflection-Transmission (RT) equations for
these models. In the case of diagonal $S$-matrix in the bulk, one of the
solutions of these equations corresponds to the Ising model with a line of
defect. We compute the reflection and transmission amplitudes for the
interaction of the Majorana fermion with the defect in two different
ways, i.e. resumming the perturbative series and implementing the boundary
conditions of the equations of motion.

\resection{Reflection-Transmission Equations}

Consider an integrable 2-d model in the bulk, defined by an action
\EQ
{\cal A}_B\,=\,\int_{-\infty}^{+\infty} dx\,\int_{-\infty}^{+\infty} dy
\, {\cal L}\left(\partial_{\mu}\phi_i,\phi_i\right) \,\, ,
\EN
where $\phi_i$ are local fields of the theory. Let us assume we have
completely solved the dynamics of the corresponding
theory in the Minkoswki space and as result, we know the mass spectrum
$\{m_a\}$ and the bulk scattering amplitudes $S_{ab}^{cd}(\beta_{ab})$
\footnote{$\beta_{ab}=
\beta_a-\beta_b$, where $\beta_i$ is the rapidity variable of the particle
$A_i$. It is related to the momenta by $p_0^{(i)}=m_i\cosh\beta_i$,
$p_1^{(i)}=m_i\sinh\beta_i$.}. The presence of a linear defect line can be
described by adding to the action of the bulk an interaction localized, say,
along the $y$-axes,
\EQ
{\cal A}\,=\,{\cal A}_B\,+\,\int_{-\infty}^{+\infty} dx \int_{-\infty}
^{+\infty} dy \, \delta(x) {\cal L}_D\left(\phi_i,\frac{d\phi_i}{dy}\right)
\,\,\, .
\label{defect}
\EN
Suppose that the introduction of this additional term does not affect the
integrability of the original model in such a way that
the action (\ref{defect}) is still supported by an infinite number of
conserved charges in involution. Since the system is invariant
along the $y$-direction (which we choose to identify with the time axis in the
Minkoswki space), the energy is one of the integrals of motions. On the other
hand, the momentum will not be generally conserved and we may have processes
with an exchange of momentum on the defect line, compatible though with
the conservation of the energy.

The conservation of the energy and of the other higher charges assures the
complete elasticity of the scattering
processes which take place on the defect line. In particular, this means
that a particle which hits the defect line with rapidity $\beta$
can only proceed forward with the same rapidity or reverses its motion
acquiring a rapidity of $-\beta$. A further effect of the interaction
with the defect line may be a change of the label of the particle
inside its multiplet of degeneracy. The interactions of the particles
$\mid \beta;i>$ with the defect line will then be described in terms of the
transmission and reflection amplitudes, denoted respectively by
$T_{ij}(\beta)$ and $R_{ij}(\beta)$ (fig.\,1). It is convenient to introduce
the operator ${\cal D}$ associated to the defect\footnote{For simplicity
we consider the case of a defect without internal degrees of freedom.}. This
can be considered as an additional particle of the theory with zero rapidity
in the entire time evolution of the system. Its commutation relations with
the creation operators $A_{i}^{\dagger}(\beta)$ relative to the asymptotic
particles are given by
\EQ
\begin{array}{lll}
A_{i}^{\dagger}(\beta)\,{\cal D} & = & R_{ij}(\beta)\,
A_{j}^{\dagger}(-\beta) \,{\cal D} +
T_{ij}(\beta) \,{\cal D} \,A_{j}^{\dagger}(\beta) \,\,\, ;\\
{\cal D}\, A_i^{\dagger}(\beta) & = & R_{ij}(-\beta)\,{\cal D}\,
A_j^{\dagger}(-\beta) +
T_{ij}(-\beta) \,A_j^{\dagger}(\beta)\, {\cal D} \,\,\, .
\end{array}
\label{algebra}
\EN
The consistency condition of this algebra requires the unitarity equations
\EQ
\begin{array}{ll}
R_{ij}(\beta)\, R_{jk}(-\beta) + T_{ij}(\beta)\,T_{jk}(-\beta) & =
\,\delta_{ik}
\,\,\, ;\\
R_{i}^{j}(\beta)\,T_{j}^{k}(-\beta) + T_{i}^{j}(\beta)\,R_{j}^{k}(-\beta) &
= \,0\,\,\,.
\end{array}
\label{unitarity}
\EN
Additional constraints emerge from the crossing relations
\EQ
\begin{array}{lll}
R_{ij}(\beta) & = & S_{\overline j l}^{k,\overline i}(2\beta)\,
R_{kl}(i\pi -\beta) \,\,\, ;\\
T_{i\bar j}(\beta) & = & T_{ij}(i\pi-\beta) \,\, .
\end{array}
\label{crossing}
\EN
The first of (\ref{crossing}) is obtained according to the argument proposed
in \cite{GZ,KF}, and we assume the validity of the second equation as in
the bulk.

Usually the presence of an infinite number of integrals of motion implies
not only the elasticity of all scattering processes but also their
complete factorization, i.e. an $n$-particle scattering amplitude can
be expressed in terms of the elementary two-body interactions \cite{ZZ}.
A crucial step for proving the factorization property of the total $S$-matrix
is to impose the associativity condition of the algebra (\ref{algebra}).
In this case it gives rise to a set of Reflection-Transmission (RT)
equations, those which are relevant to the following considerations
shown in fig.\,2.
The first of them (fig.\,2.a) coincides with the well-known boundary
equations analysed in \cite{Cherednik,Sklyanin},
\EQ
S_{ac}^{ef}(\beta_1-\beta_2) R_{fg}(\beta_2) S_{ge}^{dh}(\beta_1+\beta_2)
R_{gb}(\beta_1)\,=\,R_{ah}(\beta_2) S_{ch}^{fe}(\beta_2+\beta_1)
R_{fg}(\beta_2) S_{eg}^{bd}(\beta_1-\beta_2)\,\,\, .
\label{refl1}
\EN
The RT equations associated to the configurations of figs. (2.b), (2.c) and
(2.d) are given respectively by
\EQ
\begin{array}{l}
S_{ac}^{lm}(\beta_1-\beta_2) \,T_{lb}(\beta_1) \,T_{md}(\beta_2)\,=\,
S_{ml}^{bd}(\beta_1-\beta_2) \,T_{cm}(\beta_2) \,T_{dl}(\beta_1)
\,\,\,;\\
S_{ac}^{fe}(\beta_1-\beta_2) \,T_{fb}(\beta_1)\,R_{ed}(\beta_2) \,=\,
R_{ce}(\beta_2)\,S_{ae}^{fd}(\beta_1+\beta_2)\,T_{fb}(\beta_1)\,\,\,;\\
S_{ac}^{fe}(\beta_1-\beta_2) \,R_{fg}(\beta_2) \,S_{ge}^{dh}(\beta_1+\beta_2)
\,T_{hb}(\beta_1)\,=\,T_{ab}(\beta_1) \,R_{cd}(\beta_2) \,\,\, . \\
\end{array}
\label{YB}
\EN
The RT equations become very restrictive once applied to models with
diagonal $S$-matrix in the bulk. In fact, whereas
eq.\,(\ref{refl1}) and the first in (\ref{YB}) are identically satisfied, the
last two equations in (\ref{YB}) become
\EQ
\begin{array}{l}
S_{ab}(\beta_a\,+\beta_b) \,=\, S_{ab}(\beta_b-\beta_a) \,\, ,\\
S_{ab}(\beta_a\,+\,\beta_b)\,S_{ab}(\beta_a\,-\,\beta_b)\,=\,1 \,\, ,
\end{array}
\label{cr}
\EN
whose solutions are only $S_{ab}(\beta) = \pm 1$. Hence, we arrive to the
conclusion that the only integrable QFT with diagonal $S$-matrix in the
bulk and factorizable scattering in the presence of the defect line
are those associated to generalized-free theory. It is an open problem
whether or not the Reflection-Transmission equations admit a non-trivial
solution in the case of non-diagonal bulk $S$-matrix.

\resection{Ising Model with a Line of Defect}
A particularly interesting theory is that associated to $S=-1$. It
corresponds to the Ising model, which is described in the bulk by the free
Majorana fermion with Lagrangian given
by\footnote{We denote with $\rho$ the couple of coordinate $(x,t)$.}
\cite{Japan,Wu}
\EQ
{\cal L}_M\,=\,\overline\Psi(\rho)\left(i\gamma^{\mu}\partial_{\mu} - m\right)
\Psi(\rho) \,\,\, .
\label{Majorana}
\EN
In the Majorana representation, given by $\gamma^0 = \sigma_2$,
$\gamma^1 = -i \sigma_1$, the fermionic field $\Psi(\rho)$ is real, i.e.
$\Psi^{\dagger}(\rho) = \Psi(\rho)$.
The Ising model with a line of defect is obtained by adding to the
Lagrangian (\ref{Majorana}) the interaction
%\cite{Bariev,McCoy,Kad,Brown,Hanke,Burk,Ko}
[16-22]
\EQ
{\cal L}_D\,=\, -g\,\delta(x) \overline\Psi(\rho)\Psi(\rho) \,\, ,
\label{defising}
\EN
where $g$ is a dimensionless coupling constant. The resulting Lagrangian
is still quadratic in the fermionic field and the beta-function associated
to the coupling constant $g$ is identically zero. As a consequence, the theory
presents a non-universal ultraviolet behaviour and the critical
exponent of the magnetization operators depends continuously on the
parameter $g$ \cite{Bariev,McCoy,Kad,Brown}.

We are interested in determining the reflection $R(\beta)$ and transmission
$T(\beta)$ amplitudes for the scattering of the fermion with the defect
line, i.e. the $S$-matrix elements between initial and final states
$u(p_i)$ and $\overline u(p_f)$ with $p_f = \pm p_i$. To this aim, let us
compute the Green-function of the fermion field using the Feynman rules
given by
\vspace{1mm}

\begin{picture}(210,25)
\put(50,8){\vector(1,0){25}}
\put(54,0){\makebox(0,0){$p$}}
\put(75,8){\line(1,0){15}}
\put(86,1){\makebox(0,0){$p'$}}
\put(100,8){\makebox(0,0)[l]{$= i (2\pi)^2\, \delta^2 (p - p') \,
\frac{\not p + m}{p^2 - m^2 + i\epsilon}$}}
\end{picture}
\begin{picture}(250,25)
\put(50,8){\vector(1,0){11}}
\put(61,8){\line(1,0){10}}
\put(75,8){\circle*{8}}
\put(79,8){\vector(1,0){11}}
\put(90,8){\line(1,0){10}}
\put(105,8){\makebox(0,0)[l]{$ = -\,i g \,2\pi\,\delta (p_0 - p_0')$}}
\end{picture}

\noindent
The self-energy entering the exact propagator is given by
\vspace{3mm}

\begin{picture}(420,50)
\put(28,25){\vector(1,0){11}}
\put(39,25){\line(1,0){10}}
\put(69,25){\oval(40,25)}
\put(69,25){\makebox(0,0){$\Sigma$}}
\put(89,25){\vector(1,0){11}}
\put(100,25){\line(1,0){10}}
\put(118,25){\makebox(0,0){$=$}}
\put(129,25){\vector(1,0){8}}
\put(137,25){\line(1,0){3}}
\put(143,25){\circle*{6}}
\put(146,25){\vector(1,0){8}}
\put(154,25){\line(1,0){3}}
\put(168,25){\makebox(0,0){+}}
\put(180,25){\vector(1,0){8}}
\put(188,25){\line(1,0){3}}
\put(194,25){\circle*{6}}
\put(197,25){\vector(1,0){8}}
\put(205,25){\line(1,0){3}}
\put(211,25){\circle*{6}}
\put(214,25){\vector(1,0){8}}
\put(222,25){\line(1,0){3}}
\put(229,25){\makebox(0,0)[l]{+}}
\put(241,25){\vector(1,0){8}}
\put(249,25){\line(1,0){3}}
\put(255,25){\circle*{6}}
\put(258,25){\vector(1,0){8}}
\put(266,25){\line(1,0){3}}
\put(272,25){\circle*{6}}
\put(275,25){\vector(1,0){8}}
\put(283,25){\line(1,0){3}}
\put(289,25){\circle*{6}}
\put(292,25){\vector(1,0){8}}
\put(300,25){\line(1,0){3}}
\put(312,25){\makebox(0,0)[l]{$+ \cdots $}}
\end{picture}

\noindent
where we have to integrate on the spatial component of the momentum
running in the internal lines. Using

\vspace{1mm}
\begin{picture}(250,50)
\put(34,25){\circle*{6}}
\put(37,25){\vector(1,0){8}}
\put(43,15){\makebox(0,0){$k$}}
\put(45,25){\line(1,0){3}}
\put(51,25){\circle*{6}}
\put(64,25){\makebox(0,0){=}}
\put(76,25){\makebox(0,0)[l]
{$ (-ig)^2\, i\,\delta(k_0 - p_0)\,\int\frac{dk^1}{2\pi}\,
\frac{\not k +m}{k^2 - m^2 + i\epsilon} \,
= \,-g^2\, \delta(k_0 - p_0)\,\frac{p_0 \gamma^0 + m}{2 \omega}
\hspace{3mm},$}}
\end{picture}

\noindent
the geometric series for $\Sigma$ can be resummed as\footnote{In comparison
with \cite{Burk}, notice that in the massless limit
our result implies a different coupling constant dependance of the Green
functions of the fermionic fields.}
\EQ
\Sigma(p_0)\,=\,2\pi\,i\, \delta(p_0-p_0')\,
\sin\chi \,\frac{\omega - i \frac{g}{2} (p_0 \gamma^0 - m)}
{\omega - i m\, \sin\chi} \,\, ,
\EN
where
\[
\omega\,=\,\sqrt{p_0^2 - m^2} \hspace{3mm},\hspace{5mm}
\sin\chi\,=\, -\,\frac{g}{1 + \frac{g^2}{4}} \,\,\, .
\]
Hence, for the transmission and reflection amplitudes defined by
\[
<\beta'\mid\beta>\,=\,2\pi \delta(\beta - \beta')\,T(\beta,g) +
2\pi \delta(\beta + \beta')\,R(\beta,g) \,\, ,
\]
we have
\begin{eqnarray}
T(\beta,g) & = & \frac{\cos\chi\,\sinh\beta}{\sinh\beta - i \sin\chi}
\,\, ,
\nonumber\\
&&
\label{rt}\\
R(\beta,g) & = & i \, \frac{\sin\chi\,\cosh\beta}
{\sinh\beta - i \sin\chi} \,\,\, .\nonumber
\end{eqnarray}
The transmission amplitude also contains the disconnected
part relative to the free motion.

An alternative derivation of the transmission and reflection amplitudes is
obtained by implementing the algebra (\ref{algebra}) on the creation operators
of the fermion field. Let $\Psi_{\pm}(x,t)$ be the solutions of the free Dirac
equation in the two intervals $x>0$ and $x<0$,
i.e.
\EQ
\Psi(\rho)\,=\,\theta(x)\,\Psi_+(x,t) \,+\,\theta(-x)\,\Psi_-(x,t)
\,\, ,
\label{full}
\EN
with the value at the origin given by $
\Psi(0,t)\,=\,\frac{1}{2}\left(\Psi_+(0,t)\,+\,\Psi_-(0,t)\right)
$. The mode expansion of the two components of the fields $\Psi_{\pm}(\rho)$
is expressed as
\begin{eqnarray}
\psi_{(\pm)}^{(1)}(x,t) & = &
\int
\frac{d\beta}{2\pi} \left[
\omega e^{\frac{\beta}{2}}\, A_{(\pm)}(\beta)\, e^{-i m(t \cosh\beta
- x\sinh\beta)} \,+\,
\overline\omega e^{\frac{\beta}{2}}\, A_{(\pm)}^{\dagger}(\beta)\,
e^{i m(t \cosh\beta - x\sinh\beta)} \right]
\label{mode}\\
\psi_{(\pm)}^{(2)}(x,t) & = & -
\int
\frac{d\beta}{2\pi} \left[
\overline\omega e^{-\frac{\beta}{2}}\, A_{(\pm)}(\beta)\, e^{-i m
(t \cosh\beta - x\sinh\beta)} \,+\,
\omega e^{-\frac{\beta}{2}}\, A_{(\pm)}^{\dagger}(\beta) \,
e^{i m (t\cosh\beta - x\sinh\beta)}\right] \,\, ,\nonumber
\end{eqnarray}
with $\omega=\exp(i\pi/4)$, $\overline\omega=\exp(-i\pi/4)$.
The operators $A_{\pm}(\beta)$ and $A^{\dagger}_{\pm}(\beta)$ satisfy the
usual anti-commutation relations. They are not independent since they are
related to each other by the conditions at $x=0$ which arise from applying
the eqs. of motion to (\ref{full})
\EQ
\begin{array}{lll}
(\psi_{+}^{(2)} - \psi_{-}^{(2)})(0,t) & = & \frac{g}{2} (\psi_{+}^{(1)} +
\psi_{-}^{(1)})(0,t) \,\,\, ;\\
(\psi_{+}^{(1)} - \psi_{-}^{(1)})(0,t) & = & \frac{g}{2} (\psi_{+}^{(2)} +
\psi_{-}^{(2)} )(0,t)\,\, ,
\end{array}
\label{bc}
\EN
i.e.
\EQ
M\,
\left(\begin{array}{c}
A_-^{\dagger}(\beta) \\
A_+^{\dagger}(-\beta)
\end{array}
\right)
\,=\,
N\,
\left(\begin{array}{c}
A_-^{\dagger}(-\beta) \\
A_+^{\dagger}(\beta)
\end{array}
\right) \,\, ,
\EN
where
\[
M\,=\,\left(
\begin{array}{ll}
\omega e^{-\frac{\beta}{2}} + \frac{g}{2} \overline\omega e^{\frac{\beta}{2}} &
\,\,\,
-\omega e^{\frac{\beta}{2}} + \frac{g}{2} \overline\omega e^{-\frac{\beta}{2}}
\\
\omega e^{-\frac{\beta}{2}} + \frac{2}{g} \overline\omega e^{\frac{\beta}{2}} &
\,\,\,
\omega e^{\frac{\beta}{2}} - \frac{2}{g} \overline\omega e^{-\frac{\beta}{2}}
\end{array}
\right)
\hspace{3mm} ;
\]
\[
N\,=\,
\left(
\begin{array}{ll}
-\omega e^{\frac{\beta}{2}} - \frac{g}{2} \overline\omega e^{-\frac{\beta}{2}}
&
\omega e^{-\frac{\beta}{2}} - \frac{g}{2} \overline\omega e^{\frac{\beta}{2}}
\\
-\omega e^{\frac{\beta}{2}} - \frac{2}{g} \overline\omega e^{-\frac{\beta}{2}}
&
-\omega e^{-\frac{\beta}{2}} + \frac{2}{g} \overline\omega e^{\frac{\beta}{2}}
\end{array}
\right)
\,\,\, .
\]
Hence,
\EQ
\left(\begin{array}{c}
A_-^{\dagger}(\beta) \\
A_+^{\dagger}(-\beta)
\end{array}
\right)
\,=\,
M^{-1}\,N\,
\left(\begin{array}{c}
A_-^{\dagger}(-\beta) \\
A_+^{\dagger}(\beta)
\end{array}
\right)
\,=\,
\left(\begin{array}{ll}
R(\beta,g) & T(\beta,g) \\
T(\beta,g) & R(\beta,g)
\end{array}
\right)
\left(\begin{array}{c}
A_-^{\dagger}(-\beta) \\
A_+^{\dagger}(\beta)
\end{array}
\right)
\EN
with $R(\beta,g)$ and $T(\beta,g)$ given in (\ref{rt}).

It is easy to see that the amplitudes (\ref{rt}) satisfy the unitarity and
crossing equations (\ref{unitarity}) and (\ref{crossing}). For negative values
of $g$ the interaction with the defect line is attractive and the theory
presents a bound state with binding energy $e_b = m\cos\chi$. Notice that
simple expressions are obtained for the partial-wave phase shifts
\[
e^{2i\delta_0}  \equiv T(\beta,g) + R(\beta,g) \,=\,
\frac{\sinh\frac{1}{2}\left(\beta + i \chi\right)}
{\sinh\frac{1}{2}\left(\beta - i \chi\right)} \,\,\, ;
\]
\[
e^{2i\delta_1} \equiv T(\beta,g) - R(\beta,g) \,=\,
\frac{\cosh\frac{1}{2}\left(\beta - i \chi\right)}
{\cosh\frac{1}{2}\left(\beta + i \chi\right)}\,\, ,
\]
where $\delta_0$ and $\delta_1$ are crossed functions of each other.
Quite interesting is also the strong-weak duality presented by the scattering
amplitudes, i.e.
\EQ
T\left(\beta,\frac{4}{g}\right)\,= \,-\,T(\beta,g)
\hspace{3mm} ,
\hspace{6mm}
R\left(\beta,\frac{4}{g}\right)\,=\,R(\beta,g) \,\,\, .
\EN
At the self-dual points $g^2=4$ the transmission amplitude vanishes
and the reflection amplitudes $R(\beta,\pm 2)$ reduce to those of the
Ising model with free ($-$) and fixed ($+$) boundary conditions, as determined
in \cite{GZ}. This can also be seen directly by analysing the resulting
boundary conditions (\ref{bc}).

\resection{Conclusion}

Integrable models with a line of defect are described, in addition to the
usual scattering processes in the bulk, by a set of reflection and
transmission amplitudes. The factorization condition of the $n$-particle
amplitude gives rise to the Reflection-Transmission equations which put severe
constraints on the possible realization of those models. In the case of
diagonal $S$-matrix in the bulk, there are only two solutions of the
RT equations, one of them corresponding to the Ising model with a line
of defect. In this paper we have determined the reflection and transmission
amplitudes relative to the interaction of the Majorana fermion with the
defect line. This formulation of the model can be very useful to
compute its correlation functions by using the form factor approach.
We hope to come back to this problem in a future publication.

\vspace{5mm}
\noindent
{\em Acknowledgements}. We are grateful to T.W. Burkhardt, J.L. Cardy,
A. Cavagna, R. Iengo and A. Schwimmer for useful discussions. Two of us,
(G.D. and G.M.) thank H. Saleur and P. Fendley for the warm hospitality at
the University of Southern California, where this work has begun.

\pagestyle{empty}

\newpage

\hs

\vspace{25mm}

{\bf Figure Captions}

\vspace{1cm}

\begin{description}
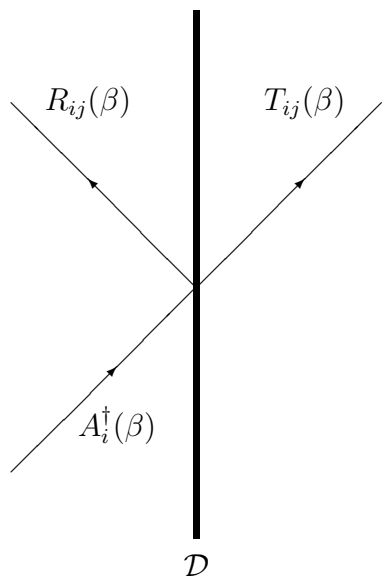
\item[ Figure 1]. Reflection and Transmission Amplitudes.
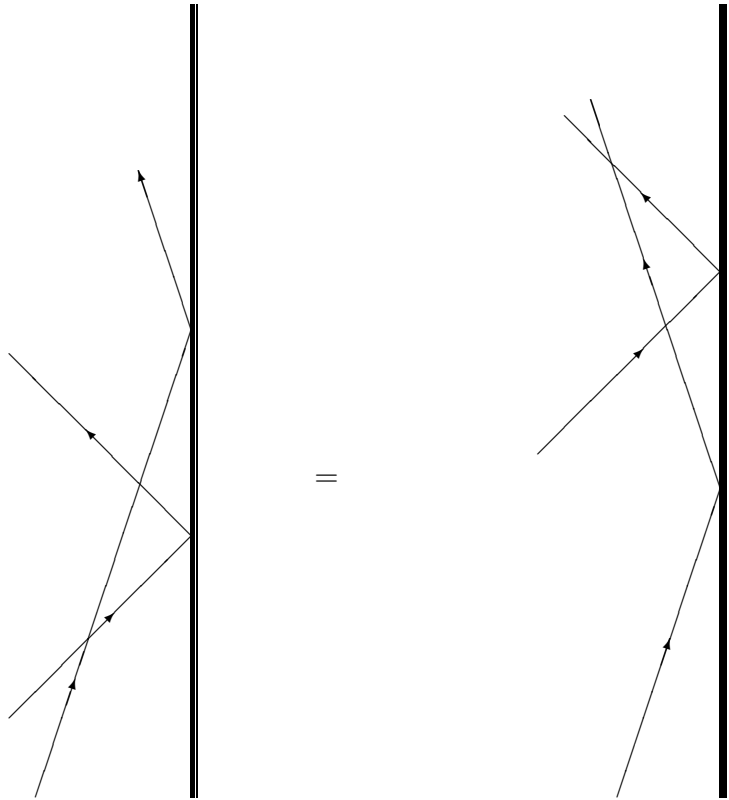
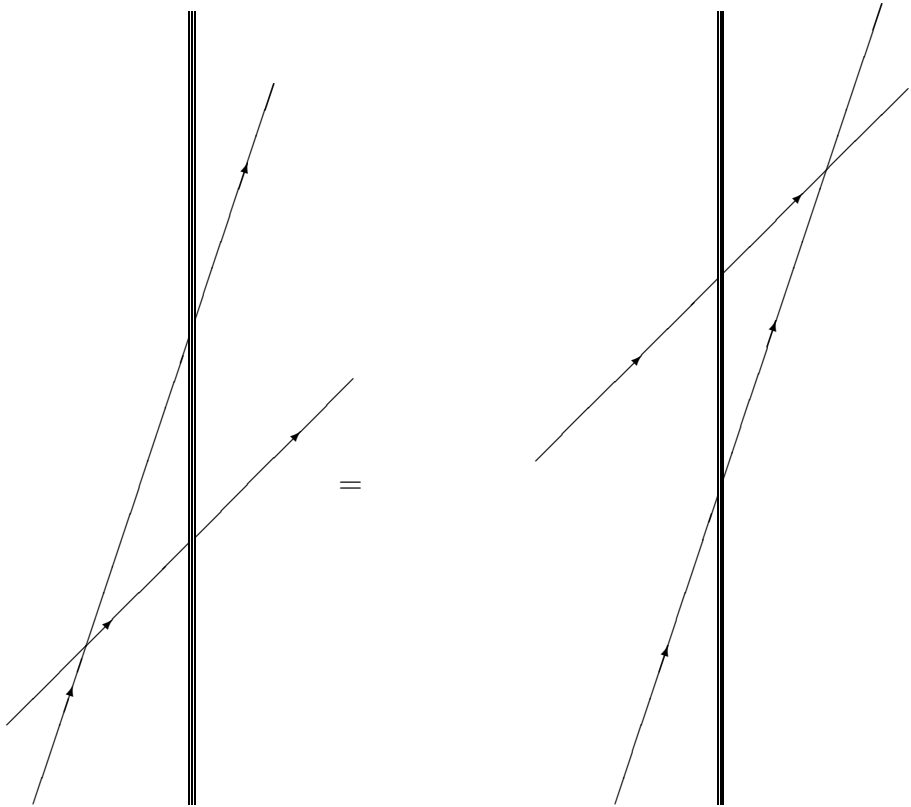
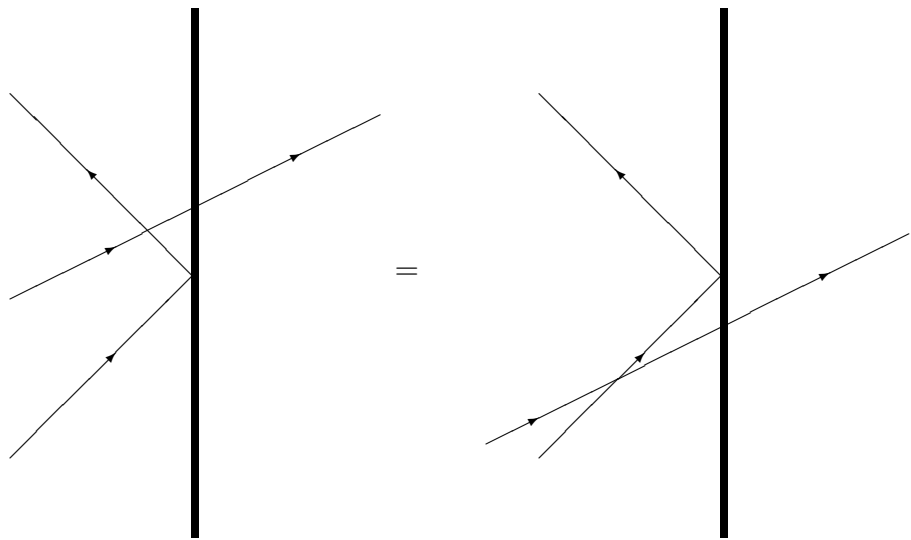
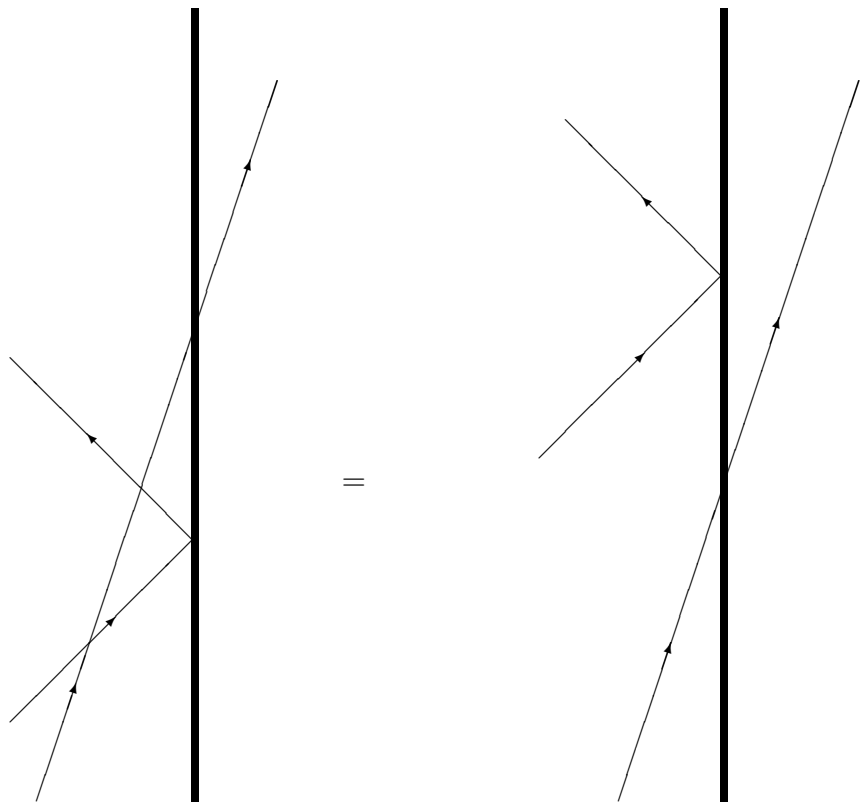
\item [ Figure 2]. Reflection-Transmission Equations.
\end{description}

\newpage
\setcounter{page}{0}

 ,

\vspace{3cm}
\begin{picture}(400,400)
\put(100,200){\begin{picture}(200,200)
\put(99,10){\line(0,1){200}}
\put(101,10){\line(0,1){200}}
\put(100,10){\line(0,1){200}}
\put(100,0){\makebox(0,0){${\cal D}$}}
\put(30,35){\vector(1,1){40}}
\put(70,75){\line(1,1){29}}
\put(70,53){\makebox(0,0){$A_i^{\dagger}(\beta)$}}
\put(99,106){\vector(-1,1){40}}
\put(59,146){\line(-1,1){29}}
\put(59,175){\makebox(0,0){$R_{ij}(\beta)$}}
\put(101,106){\vector(1,1){40}}
\put(141,146){\line(1,1){29}}
\put(141,175){\makebox(0,0){$T_{ij}(\beta)$}}
\end{picture}}
\end{picture}

\vspace{1mm}
\begin{center}
Figure 1
\end{center}

\newpage
\begin{picture}(300,500)
\put(50,100){\begin{picture}(300,400)
\put(99,0){\line(0,1){300}}
\put(101,0){\line(0,1){300}}
\put(100,0){\line(0,1){300}}
\put(30,30){\vector(1,1){40}}
\put(70,70){\line(1,1){29}}
\put(99,99){\vector(-1,1){40}}
\put(59,139){\line(-1,1){29}}
\put(40,0){\vector(1,3){15}}
\put(55,45){\line(1,3){5}}
\put(60,60){\line(1,3){39}}
\put(99,177){\vector(-1,3){20}}
\put(150,120){\makebox(0,0){$=$}}
\put(299,0){\line(0,1){300}}
\put(300,0){\line(0,1){300}}
\put(301,0){\line(0,1){300}}
\put(230,130){\vector(1,1){40}}
\put(270,170){\line(1,1){29}}
\put(299,199){\vector(-1,1){30}}
\put(269,229){\line(-1,1){29}}
\put(260,0){\vector(1,3){20}}
\put(280,60){\line(1,3){19}}
\put(299,117){\vector(-1,3){29}}
\put(270,204){\line(-1,3){20}}
\end{picture}}
\end{picture}
\vspace{1mm}
\begin{center}
\hspace{15mm} Figure 2.a
\end{center}
%Reflection

\newpage
\begin{picture}(400,500)
\put(0,100){\begin{picture}(400,400)
\put(99,0){\line(0,1){300}}
\put(101,0){\line(0,1){300}}
\put(100,0){\line(0,1){300}}
\put(30,30){\vector(1,1){40}}
\put(70,70){\line(1,1){29}}
\put(101,101){\vector(1,1){40}}
\put(141,141){\line(1,1){20}}
\put(40,0){\vector(1,3){15}}
\put(55,45){\line(1,3){5}}
\put(60,60){\line(1,3){39}}
\put(101,183){\vector(1,3){20}}
\put(121,243){\line(1,3){10}}
\put(160,120){\makebox(0,0){$=$}}
\put(299,0){\line(0,1){300}}
\put(300,0){\line(0,1){300}}
\put(301,0){\line(0,1){300}}
\put(230,130){\vector(1,1){40}}
\put(270,170){\line(1,1){29}}
\put(301,201){\vector(1,1){30}}
\put(331,231){\line(1,1){40}}
\put(260,0){\vector(1,3){20}}
\put(280,60){\line(1,3){19}}
\put(301,123){\vector(1,3){20}}
\put(321,183){\line(1,3){40}}
\end{picture}}
\end{picture}
\vspace{1mm}
\begin{center}
\hspace{5mm} Figure 2.b
\end{center}

\newpage
\begin{picture}(400,400)
\put(30,100){\begin{picture}(200,300)
\put(99,0){\line(0,1){200}}
\put(101,0){\line(0,1){200}}
\put(100,0){\line(0,1){200}}
\put(30,30){\vector(1,1){40}}
\put(70,70){\line(1,1){29}}
\put(99,99){\vector(-1,1){40}}
\put(59,139){\line(-1,1){29}}
\put(30,90){\vector(2,1){40}}
\put(70,110){\line(2,1){50}}
\put(120,135){\vector(2,1){20}}
\put(140,145){\line(2,1){30}}
\put(180,100){\makebox(0,0){$=$}}
\put(300,0){\line(0,1){200}}
\put(299,0){\line(0,1){200}}
\put(301,0){\line(0,1){200}}
\put(230,30){\vector(1,1){40}}
\put(270,70){\line(1,1){29}}
\put(299,99){\vector(-1,1){40}}
\put(259,139){\line(-1,1){29}}
\put(210,35){\vector(2,1){20}}
\put(230,45){\line(2,1){40}}
\put(270,65){\line(2,1){50}}
\put(320,90){\vector(2,1){20}}
\put(340,100){\line(2,1){30}}
\end{picture}}
\end{picture}

\vspace{1mm}
\begin{center}
\hspace{15mm} Figure 2.c
\end{center}

\newpage
\begin{picture}(300,500)
\put(30,100){\begin{picture}(300,400)
\put(99,0){\line(0,1){300}}
\put(101,0){\line(0,1){300}}
\put(100,0){\line(0,1){300}}
\put(30,30){\vector(1,1){40}}
\put(70,70){\line(1,1){29}}
\put(99,99){\vector(-1,1){40}}
\put(59,139){\line(-1,1){29}}
\put(40,0){\vector(1,3){15}}
\put(55,45){\line(1,3){5}}
\put(60,60){\line(1,3){39}}
\put(101,183){\vector(1,3){20}}
\put(121,243){\line(1,3){10}}
\put(160,120){\makebox(0,0){$=$}}
\put(299,0){\line(0,1){300}}
\put(300,0){\line(0,1){300}}
\put(301,0){\line(0,1){300}}
\put(230,130){\vector(1,1){40}}
\put(270,170){\line(1,1){29}}
\put(299,199){\vector(-1,1){30}}
\put(269,229){\line(-1,1){29}}
\put(260,0){\vector(1,3){20}}
\put(280,60){\line(1,3){19}}
\put(301,123){\vector(1,3){20}}
\put(321,183){\line(1,3){30}}
\end{picture}}
\end{picture}
\vspace{1mm}
\begin{center}
\hspace{5mm} Figure 2.d
\end{center}
\end{document}